\documentclass[conference]{IEEEtran}
\usepackage{cite}
\usepackage{amsmath,amssymb,amsfonts}
\usepackage{algorithmic}
\usepackage{graphicx}
\usepackage{textcomp}
\usepackage{xcolor}
\usepackage{comment}
\usepackage{hyperref}
\usepackage{enumitem}
\usepackage{tabularx}
\usepackage[normalem]{ulem} 
\usepackage{balance}
\usepackage[none]{hyphenat}
\usepackage{fancyhdr}
\begin{document}
\setlist[itemize,1]{leftmargin=\dimexpr 25pt \relax}   
\setlist[itemize,2]{leftmargin=\dimexpr 25pt \relax}   
\setlist[itemize,3]{leftmargin=\dimexpr 25pt \relax}   
\title{L-PRISMA: An Extension of PRISMA in the Era of Generative Artificial Intelligence (GenAI)}

\author{\IEEEauthorblockN{Samar Shailendra, Rajan Kadel, Aakanksha Sharma, Islam Mohammad Tahidul, Urvashi Rahul Saxena}
\IEEEauthorblockA{\textit{School of IT \& Engineering (SITE), Melbourne Institute of Technology}, 
Melbourne, Australia.\\
s\_samar@ieee.org, rkadel@mit.edu.au, aasharma@mit.edu.au, mtislam@mit.edu.au, usaxena@mit.edu.au
}
}

\maketitle
\thispagestyle{fancy} 
\begin{abstract}
The Preferred Reporting Items for Systematic Reviews and Meta-Analyses (PRISMA) framework provides a rigorous foundation for evidence synthesis, yet the manual processes of data extraction and literature screening remain time-consuming and restrictive. Recent advances in Generative Artificial Intelligence (GenAI), particularly large language models (LLMs), offer opportunities to automate and scale these tasks, thereby improving time and efficiency. However, reproducibility, transparency, and auditability, the core PRISMA principles, are being challenged by the inherent non-determinism of LLMs and the risks of hallucination and bias amplification. To address these limitations, this study integrates human-led synthesis with a GenAI-assisted statistical pre-screening step. Human oversight ensures scientific validity and transparency, while the deterministic nature of the statistical layer enhances reproducibility. The proposed approach systematically enhances PRISMA guidelines, providing a responsible pathway for incorporating GenAI into systematic review workflows.


\end{abstract}


\section{Introduction} \label{sec:introduction}

Traditionally, literature surveys have played a crucial role in the research ecosystem. They serve as an essential means to compile, evaluate, and disseminate the vast amount of scientific work produced across disciplines. The survey papers based on such literature surveys have found increasing importance in the research community. Given the rapid pace at which new studies get published, survey papers play a pivotal role in distilling complex and scattered findings into coherent, consolidated and accessible narratives guiding both current research and future investigations. However, with increasing reliance on survey papers, the need was realised to standardise the process of literature survey, ensuring the accountability, reproducibility, and scientific rigour of these papers. 

\subsection{PRISMA Background}
In 2009, recognising the need for a standardised, transparent, and rigourous approach to conduct literature reviews, a systematic literature survey framework named \textit{Preferred Reporting Items for Systematic Reviews and Meta-Analyses (PRISMA)} was proposed~\cite{Moher2009}. An updated version of this PRISMA framework, which is the cornerstone for systematic reviews, was released in 2020~\cite{Pagen71}. This new version refined the original guidelines to make them more comprehensive and methodologically rigourous, improving the clarity and reproducibility of research reporting.

In spite of its systematic approach, PRISMA is not without challenges~\cite{ioannidis2016mass, dariusz2025}. Due to exponential growth in published research, even the well-crafted search strategies return thousands of records, making comprehensive screening both time-consuming and resource-intensive. To circumvent this intractable number and keep the process humanly manageable, search queries are frequently restricted by keywords, filters, time windows or lexical similarity tools to filter the results. However, this introduces the risk of inadvertently excluding relevant studies that do not match the chosen criteria or keywords. Moreover, the manual effort required for title and abstract screening, followed by full-text review, places a heavy cognitive burden on researchers and increases the likelihood of inconsistencies or human error.  As a result, despite PRISMA’s rigour and structure, systematic reviews remain constrained by limitations in search breadth, imperfect human filtering, and the practical impossibility of exhaustively reviewing very large volumes of literature.

In the recent past, there has been humongous growth in Natural Language Processing (NLP), enabling computers to better understand, interpret, and analyse human language. These advancements, particularly the introduction of Transformer architectures~\cite{vaswani2017}, have dramatically improved the ability of machines to process large volumes of unstructured text with contextual awareness. Fuelled by these technical advancements, Generative Artificial Intelligence (GenAI) has opened up new possibilities for understanding language and producing coherent, contextually relevant, and human-like text. 

Large Language Models (LLMs) such as Generative Pre-trained Transformer (GPT)~\cite{Radford2018ImprovingLU,yenduri2024}, Claude~\cite{Bai2022ConstitutionalAH}, and Gemini~\cite{geminiteam2025geminifamilyhighlycapable}, alongside Transformer-based tools like BERT, combine these capabilities to both comprehend and generate information at scale. These models can extract key insights, summarise findings, and adapt to domain-specific requirements, demonstrating transformative potential across research-intensive disciplines. The researchers can potentially automate and accelerate the laborious processes of screening, and data extraction using LLMs, thereby broadening the scope of systematic reviews. However, the integration of these powerful yet opaque tools introduces a new set of fundamental challenges that directly conflict with PRISMA’s core tenets of transparency, reproducibility, ethical usage, and auditability \cite{dariusz2025, eacersall2024}.

\subsection{Issues in PRISMA Framework in the Age of GenAI}
GenAI is successfully able to handle the challenges of scale and manual effort however that has constrained the PRISMA framework~\cite{ioannidis2016mass, dariusz2025}. The first major challenge lies in reproducibility. A traditional PRISMA search strategy, built on specific keywords and database queries, is designed to be precisely documented and replicated by other researchers. However, GenAI-driven discovery is often non-deterministic. The results can vary based on the specific model used, its version, hallucinate, and subtle variations in prompting. This ``black box'' nature makes an AI-assisted search process exceptionally difficult to report in a way that guarantees another researcher to achieve the same outcome. To circumvent this we have suggested a GenAI assisted statistical approach, where we do the pre-screening using statistical approach which keeps the outcome deterministic and reproducible however, does not strain the researchers with deep understanding of the statistics.

\begin{table*}[htbp]
\centering
\caption{Issues in Applying PRISMA Framework in the Age of GenAI.}
\label{tab:PRISMA_GenAI_Issues}
\begin{tabular}{|p{2.5cm}|p{5cm}|p{9cm}|}
\hline
\textbf{Area} & \textbf{Issue} & \textbf{Explanation} \\
\hline
Search Strategy  Transparency & AI-assisted searches may not be reproducible~\cite{gundersen2025unreasonable,haibe2020transparency} & AI-generated search strategies (e.g., from ChatGPT) may not provide full transparency or consistency, making it difficult to replicate systematic searches. \\
\hline
Selection Bias & GenAI may introduce or amplify bias~\cite{schwartz2022towards,lloyd2018bias} & LLMs are trained on existing data and may prioritise certain sources or viewpoints, potentially skewing literature selection. \\
\hline
Screening and Eligibility & Over-reliance on GenAI for inclusion/exclusion decisions & Delegating decision-making to AI tools can compromise human judgment and contextual understanding, particularly in nuanced inclusion criteria. \\
\hline
Data Extraction & AI automation may misinterpret nuanced data~\cite{adel2025genAI} & AI can help extract structured information, but it may struggle with context-specific or discipline-specific data interpretation. \\
\hline
Critical Appraisal & Lack of rigourous quality assessment by AI~\cite{ZYBACZYNSKA2024106} & AI lacks the critical thinking needed for nuanced evaluation of study quality, risking the inclusion of low-quality or biased studies. \\
\hline
Updating and Living Reviews & Automated updates may bypass human oversight~\cite{bolanos_artificial_2024} & AI can assist with “living systematic reviews,” but without rigourous validation, automatic updates might integrate unverified or low-quality sources. \\
\hline
Plagiarism \& Redundancy Risk & AI-generated content may mirror training data~\cite{9651895}& LLM-generated summaries or interpretations might unintentionally reproduce or closely paraphrase existing literature without proper attribution. \\
\hline
PRISMA Flow Diagram Generation & Automated tools might misrepresent workflow~\cite{krol_responsible_2025}& Some GenAI-based tools might produce incomplete or overly simplified diagrams, failing to accurately reflect the review process. \\
\hline
Ethics and Authorship & Unclear role of AI in authorship~\cite{andrade-hidalgo_exploring_2025, knight_understanding_2025} & Lack of clarity on how to credit GenAI contributions in systematic reviews poses ethical and scholarly dilemmas. \\
\hline
Overconfidence in AI Output & Hallucinations or inaccurate citations~\cite{ji2023, rawte2023} & GenAI can generate plausible but fabricated references or misattribute findings, leading to misinformation in systematic reviews. \\
\hline
AI-augmented systematic reviews & Manual screening and data extraction are time-consuming and non-scalable~\cite{adel2025genAI} & Ensuring computational efficiency while preserving the epistemic standards of transparency, reproducibility, and rigour thereby avoiding over-reliance on opaque models while leveraging their strengths to reduce workload.\\
\hline
Reproducibility in AI workflows & LLM outputs are non-deterministic, making replication of results challenging~\cite{bolanos2024AIreview}. & Identifies reproducibility gaps introduced by GenAI and stresses the need for deterministic methods or reporting protocols to ensure transparency.\\
\hline
Hybrid frameworks & Purely AI-driven pipelines risk errors and lack robustness~\cite{ge2024health} & Proposes combining statistical pre-screening with AI methods to balance efficiency with reproducibility, ensuring methodological soundness.\\
\hline
Automation of Systematic Reviews & Manual literature reviews are time-consuming, resource-intensive, and often hindered by the rapid growth of publications~\cite{torre2023automate} & AI-driven approaches can automate key stages of systematic reviews, such as document retrieval, screening, and data extraction thereby enhancing scalability while maintaining methodological rigour.\\
\hline
Transparency and auditability & Lack of traceability in AI-augmented PRISMA processes~\cite{li2025AMA} & Advocates for structured audit trails, transparent documentation, and adherence to established standards (e.g., PRISMA) to preserve scientific integrity.\\
\hline
\end{tabular}
\end{table*}

The automation of screening and data extraction raises concerns about reliability and scientific rigour. LLMs may misinterpret nuanced scientific arguments, fail to distinguish high-quality studies from flawed ones, or generate ``hallucinations''—confident but entirely fabricated citations and summaries~\cite{ji2023,rawte2023}. This shifts the researcher's burden from manual screening to the equally demanding task of meticulously verifying the AI's output. Issues of authorship and intellectual responsibility for the final reported findings are further complicated with a risk of bias. Hence, we propose to review the most relevant literature following the traditional human based, while strategically incorporating GenAI-based searches to identify potentially overlooked studies, thereby maximising comprehensiveness without sacrificing rigour. 

The study by Gundersen et al.~\cite{gundersen2025unreasonable} offers convincing proof of the efficacy of open science procedures in AI, showing that reproducibility is both a scientific and a technical prerequisite for maintaining accountability and facilitating cumulative research. Their initial findings are in line with the difficulties encountered in GenAI-driven literature reviews, where the non-deterministic nature of LLMs compromises reproducibility. The research proposed by Haibe-Kains et al.~\cite{haibe2020transparency} serves as a foundation for understanding the limitations of non-deterministic AI methods (LLMs) in systematic reviews and PRISMA-like workflows. This highlights the significance of open and consistent processes, especially when incorporating AI into frameworks for evidence synthesis like PRISMA. In addition, Schwartz et al.~\cite{schwartz2022towards} provide a systematic approach for recognizing and controlling bias in AI, along with useful tactics to reduce risks associated with automated screening and data extraction. Their viewpoint is in line with Lloyd's~\cite{lloyd2018bias}, who cautions that AI systems have the potential to reinforce preexisting biases through feedback loops, skewing the breadth and interpretation of research. Together, these results show that although GenAI offers chances to make systematic reviews more scalable, its incorporation needs to be done carefully, with protections for reproducibility, transparency, and bias reduction to maintain scientific integrity. Further, to ensure transparency, validity, and reproducibility across the design, collection, and interpretation stages of AI-augmented systematic reviews, Malik and Terzidis~\cite{malik2025hybrid} suggested a hybrid approach that strikes a balance between computational efficiency and epistemic rigour. 

Recent developments in the integration of AI into workflows for systematic reviews have begun to resolve the scalability issues associated with evidence synthesis.  Numerous studies demonstrate how GenAI technologies can automate data extraction and filtering, lowering manual labour costs and increasing the scope of searchable literature~\cite{adel2025genAI}.  However, because LLM-driven processes produce non-deterministic outputs that make transparency and independent validation more difficult, repeatability issues still exist~\cite{bolanos2024AIreview}.  Concurrently, studies on hybrid frameworks emphasise how important it is to combine deterministic statistical techniques with AI-based automation in order to address replicability concerns and guarantee methodological robustness~\cite{ge2024health}.  Furthermore, research on AI-assisted review pipelines highlights the significance of bias mitigation and detection techniques, cautioning that a blind dependence on machine-generated outputs can sustain persistent errors in the synthesis of evidence~\cite{torre2023automate}. In addition, to match AI-augmented procedures with well-established protocols like PRISMA and preserve scientific rigour and credibility, emerging systems also provide public reporting requirements and organised audit trails~\cite{li2025AMA}.  The potential of GenAI in speeding up systematic reviews is supported by this body of work, which also emphasises the critical need for protections that maintain reproducibility, transparency, and bias control. Table~\ref{tab:PRISMA_GenAI_Issues} summarises the issues in applying PRISMA framework in the era of GenAI. 

\subsection{Motivations and Contributions}
Based upon the limitation of PRISMA framework as discussed in the previous section, we propose to refine PRISMA framework to include GenAI based tools without compromising the scientific rigour and transparency that PRISMA has established. This paper provides a clear pathway for leveraging these powerful technologies to enhance the quality of systematic reviews. The primary contributions of this paper are:

\textbf{\textit{An Enhanced PRISMA Framework:}} A modified PRISMA checklist and an updated flow diagram that explicitly incorporate stages for GenAI application.

\textbf{\textit{Systematic GenAI Integration:}} A set of statistical approaches (using GenAI assistance) for systematic reviews including guidelines for reporting to ensure output verification. This will help reader to reproduce the systematic search results.

\textbf{\textit{A Hybrid Review Approach:}} A GenAI augmented review methodology ensuring the reliability of the traditional human-driven review as the core component, while systematically employing GenAI as a supplementary tool for discovery to broaden the search scope and identify literature that might be missed by conventional keyword-based strategies.

The outline of the paper is as follows: Section~\ref{sec:survey} provides a brief review of related literature. Section~\ref{sec:l-prisma} provides the details about the LLM augmented - Preferred Reporting Items for Systematic Reviews and Meta-Analyses (L-PRISMA) framework followed by use case example with analysis in Section~\ref{UCE}. Finally, Section~\ref{sec:conclusions} presents the concluding remarks and the future directions.

\section{Literature Survey}\label{sec:survey}
This section presents recent work in the area of systematic review, different tools being proposed for literature survey and different tools for systematic review.  

Toth et al.~\cite{Toth2024} map and evaluate the landscape of automation tools available to support systematic reviews, categorising them across PRISMA workflow stages (searching, screening, data extraction, synthesis, and reporting), and provide details on how they can be used to streamline evidence synthesis. However, this work has not utilised the true potential of modern-day NLP tools and LLMs to enhance both the width and depth of the literature survey.  In~\cite{Schmidt2021}, the authors have proposed a semi-automated approach to literature surveys; however, they do not outline any systematic criteria for selecting the studies that should be subject to human review. The absence of a defined selection framework raises the risk of unintentionally excluding relevant works or incorporating irrelevant ones. This necessitates a principled mechanism to guide the inclusion of literature for systematic and human review. 

Kolaski et al.~\cite{Kolaski2023}, along with PRISMA, have evaluated and compared other existing standards (AMSTAR-2, ROBIS, and GRADE). They have summarised them into a concise guide with each tool's capabilities and listed best-practice resources to improve the conduct and reporting of systematic reviews. However, this must be noted that unlike PRISMA, A Measurement Tool to Assess systematic Reviews (AMSTAR-2)~\cite{Shea2017AMSTAR2} is primarily effective for medicine and healthcare-related fields, while A Risk of Bias Assessment Tool for Systematic Reviews (ROBIS)~\cite{Whiting2016ROBIS} and Grading of Recommendations Assessment, Development, and Evaluation (GRADE)~\cite{Guyatte081903} handle the risk of bias and quality of systematic review, respectively. Furthermore, their work lacks an operational or validated proposal/framework for the systematic review, especially adaptable to the new and emerging era of GenAI.

In~\cite{bahor2021development}, authors have introduced Systematic Review Facility (SyRF), a web-based platform tailored to support preclinical systematic reviews, offering functionalities for screening, data management, and collaborative workflows; however, this work is limited only to preclinical studies mainly in the domain of medicine and does not generalise to other domains, restricting its broader applicability. There are many literature survey tools (e.g. Rayyan~\cite{rayyan_evaluation}, Covidence~\cite{Covidencecowie2022web}, RevMan~\cite{RevMan_wang2019recommended}, ASReview~\cite{ASReview_quan2024utilizing}) are also proposed in the literature. Due to the study's limitation to two systematic literature reviews, low-to-moderate specificity values, and dependence on the quality of its training set, the findings might not apply to other tools or reviews. The classifier is used to detect Randomised Controlled Trials (RCTs)~\cite{Covidencecowie2022web} when a sizeable portion of the papers it identifies as Possible RCT are not actually RCTs. The user still has to manually search through a significant number of irrelevant references in order to find the real RCTs. Because of the paper's limited statistical tools, lack of user-friendliness, and graphic alteration options, its findings are only applicable to consumers~\cite{RevMan_wang2019recommended}. The computational demands of more sophisticated ASReview models such as Doc2Vec and SBERT, might lead to longer processing times, particularly when working with large datasets~\cite{ASReview_quan2024utilizing}. While these tools currently use AI to refine and summarise the literature, they still lack the systematic approach as provided by the PRISMA framework. Moreover, the limitation of AI training, complexity of statistical tools, longer processing times for unexpected size of dataset, makes it complex for the efficient result. Recently, Teo~\cite{Susnjak2023PRISMA} have proposed a domain-specific fine-tuned LLM-based extension for PRISMA. However, this requires fine-tuning the LLMs on the domain-specific literature, which often is resource-intensive, time-consuming and requires specialised skill to fine-tune the model. This defeats the ease of using the model and makes the overall approach accessible only to researchers with deep knowledge of LLMs, defeating the simplicity and availability of the framework for everyone. There are also certain LLMs (such as Perplexity~\cite{PerplexityAI}) which claim to be tailored for research and optimised not to hallucinate or provide non-factual content by cross-referencing them to the actual reference.

A review of current literature suggests that traditional literature review methodologies often lack in research selection criteria that could lead to the exclusion of the necessary information. Although semi-automated methods offer some improvement, they still necessitate a thorough manual examination of extraneous content, while many platforms are domain-specific e.g., medicine, which restricts their broad usage. Furthermore, the structured and comprehensive framework required for a thorough and repeatable assessment is absent in the current AI era. This demands a more comprehensive framework, which offers a methodical, principled approach, directing the systematic inclusion of literature, ensuring that the review is thorough, objective, and reproducible to accommodate new and developing technologies like GenAI.

\section{L-PRISMA Framework} \label{sec:l-prisma}

PRISMA has been by far the most successful framework for systematic literature review. Based upon the extensive literature survey as in Section~\ref{sec:survey}  as well as our own experiences~\cite{data6020018,10629211}, we found certain limitations in the PRISMA framework, which if addressed, can significantly enhance the quality and depth of the systematic literature review. Hence, it is essential that PRISMA formally adopt GenAI as an integral part of it and include definitive guidelines for systematic literature review using GenAI. 

\begin{figure}
    \centering
    \includegraphics[scale=0.6]{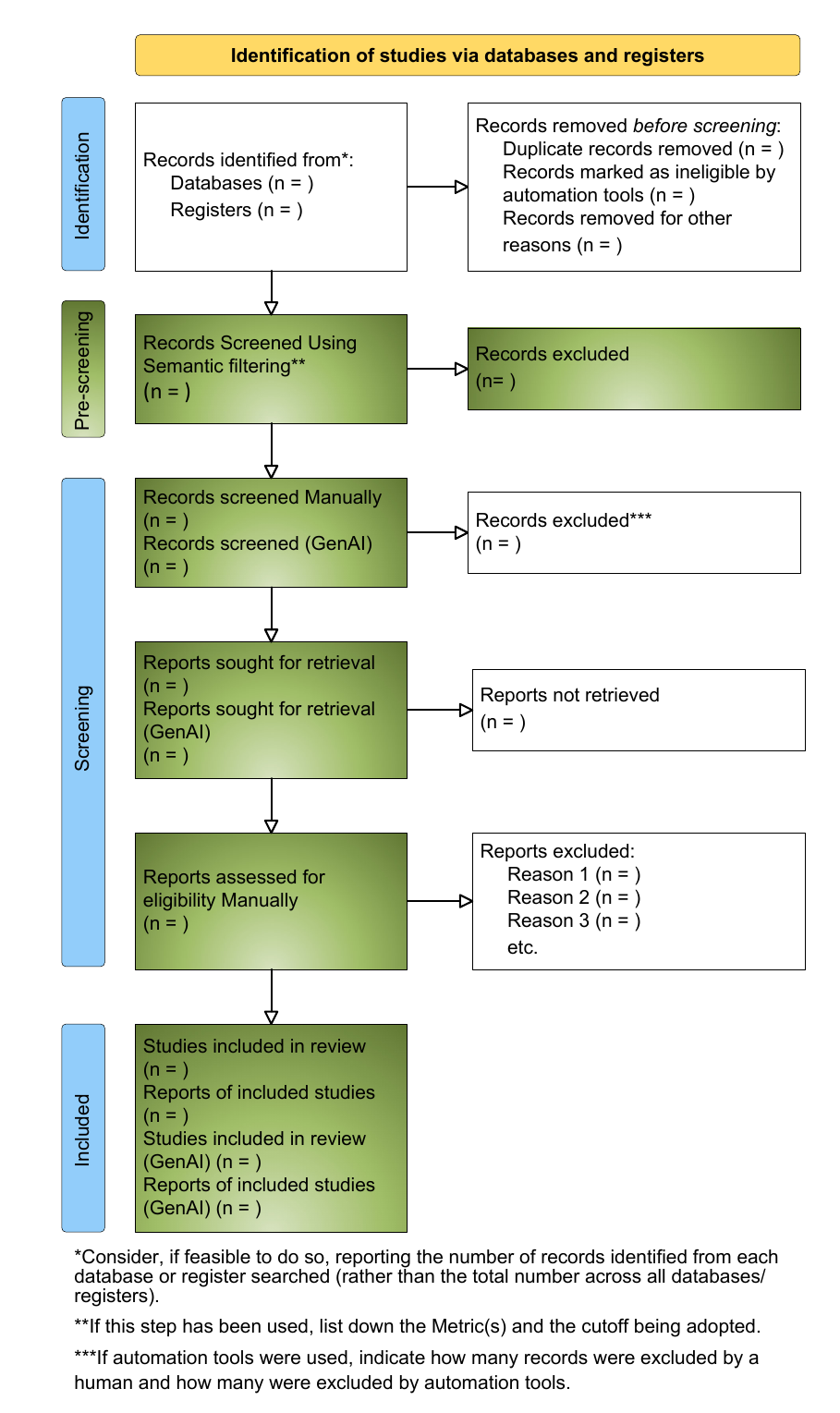}
    \caption{L-PRISMA flow diagram.}
    \label{fig:l-prisma}
\end{figure}

In this paper, we propose L-PRISMA, an extension of PRISMA framework in the era of GenAI for systematic literature review. Our framework leverages semantic filtering, transformer-based models and LLM-based tools for literature summarisation. It also provides the statistical guidelines to further refine and filter out the tool-based search results. Furthermore, unlike other proposals in literature, it does so in resource-constrained way and is easily usable by the non IT-savvy researchers, in line with the core spirit of PRISMA of being lightweight and usable by all. It must also be noted that since LLMs have the tendency to hallucinate~\cite{ji2023, rawte2023} and generate false results confidently, caution needs to be exercised to review all outcomes by a human during the process. 

The L-PRISMA framework (Fig.~\ref{fig:l-prisma})  proposes following updates over PRISMA: (i) Adding a new pre-screening phase for semantic filtering, (ii) Updating Screening phase to report GenAI records, (iii) Update Included phase to report GenAI reports and studies. During the pre-screening phase the searched records are filtered using the semantic filtering at the same time some of the categorised records during this phase are summarised using GenAI. The details of these phases are being discussed in below subsections.

\subsection {Pre-Screening Phase}
Unlike PRISMA, which suggests the use of tools for record filtering during its screening phase, L-PRISMA adds the semantic filtering as pre-screening phase which can be followed by the screening phase as described by PRISMA using the lexical similarity tools. During pre-screening, we are likely to get large number of papers searched by the search criteria. It must be noted here that narrowing the search criterion by restricting the key words or regular expression can severely limit the search and omit some important piece of work.  During this phase we use the the semantic similarity score for sorting out the relevant papers. For calculating this similarity score, the authors defines a statement which describes their intent for review and a semantic similarity score is calculated. Subsequently the histogram of the obtained score is plotted. We can approximate similarity distribution as a combination of two (or more) overlapping groups. Without loss of generality, this distribution can be represented as a mixture of two (or more) distributions: one corresponding to highly relevant articles and the other to weakly relevant (or non-relevant) articles, as formalised in Eq.~\eqref{eq:mixture-scores} based on similarity search~\cite{kanoulas2009modeling,kanoulas2010score,stephen2007}. 

\begin{equation}
\begin{aligned}
  p(s) &= \pi_{H}\,p_{H}(s) 
         + \pi_{L}\,p_{L}(s), 
         && s \in [0,1] \\
  \pi_{H} &+ \pi_{L} = 1
\end{aligned}
\label{eq:mixture-scores}
\end{equation}

where, \\
\indent $p(s)$ denotes the overall similarity score distribution, \\
\indent $p_{H}$ denotes the high relevance score distribution,\\
\indent $p_{L}$ denotes the low relevance score distribution, and \\
\indent $\pi_{H}, \pi_{L} \geq 0$ are the mixture weights.

\noindent Subsequently, 
quartile points or any other statistical boundary rule can be used to decide the decision boundaries.


This stage will capture if there are certain records being excluded and records marked for human screening during the screening phase. We suggest that this problem can be solved using LLMs. The authors do not necessarily need to be aware of any statistical tools. This allows the user to take informed decision how many papers to include for human review and how many to be included for the GenAI review or to be excluded. A use case example to demonstrate the approach is presented in the next section.  


\subsection{Screening Phase}
This phase in L-PRISMA is same as the standard PRISMA framework except this reports the number of the records to be screened manually along with the records to be screened and summarised using GenAI. The number of records which are to be excluded based on Manual screening should also be reported. 

\subsection{Included Phase}
Traditionally, PRISMA implicitly assumes that all studies included in the review are undertaken by human. While human review is important and the corner stone of the literature review, there is a physical limit over the number of reviews taken by humans. This also risk that the studies which are quite relevant has been omitted from the review. To alleviate this problem, we recommend to divide the studies into two parts, the reports which are of high relevance (e.g. the ones with high similarity score), should be summarised by the human, however, the reports with low/weak relevance can be summarised by GenAI with proper prompt so that the outcome is aligned with the overall review structure. This will ensure that the system review does not discard or miss important aspects of the literature. It must also be noted that GenAI outcomes are not perfect so they should always be human moderated and checked for their consistency.

\section{Use Case Example} \label{UCE}
This section provides the results from one of the practical use cases scenario and explains how the additional phases of L-PRISMA can help with the systematic review process.  

\subsection{Phase-1: Identification}
We carried out our search over two databases - IEEE and ACM. The search queries used and the number of records returned are listed in the Table \ref{tab:PRISMA_query}. The original intent is to find out how GenAI tools for text similarity are being used in the domain of Education. There are two types of queries one with the ``Education'' word being used in the search query and another a search without the specific Education domain. Clearly, if the record search is limited by the studies in the Education domain, it has more tractable search in terms of number and highly relevant to the Education domain, however, it easy to visualise that critical literature related to the text similarity and GenAI will not be included. To keep the number human tractable, the researcher may keep the search the restrictive and risk losing significant relevant literature. However, L-PRISMA is not constrained by this syndrome, and recommends to include all the searched records. At the same time, we also acknowledge the broader search might include some work which may not be relevant to the original literature survey. To alleviate this problem a pre-screening phase, as discussed in next section, is being proposed which using GenAI capabilities, statistically eliminate the noisy literature.

\begin{table*}[htbp]
\centering
\caption{Search query and number of results from different database (Searched on 01/Aug/2025)}
\label{tab:PRISMA_query}
\begin{tabular}{|p{1.2cm}|p{11cm}|p{1.8cm}|p{2cm}|}
\hline
\textbf{Databases} & \textbf{Search Query} & \textbf{Search Scope}  & \textbf{No. of Records} \\
\hline
        IEEE & (((``semantic'' OR ``similarity''):Abstract OR (``semantic'' OR ``similarity''):``Document Title'') AND  
        ((``natural language processing'' OR NLP OR ``Generative Artificial Intelligence'' OR GenAI OR ``Gen AI''):Abstract OR (``natural language processing'' OR NLP  OR ``Generative Artificial Intelligence'' OR GenAI OR ``Gen AI''):``Document Title'') AND ((``Education''):Abstract OR (``Education''):``Document Title'')) & With Educational Domain Constraint & 24 \\
         \hline
         ACM & ((Title:(semantic OR similarity) OR Abstract:(semantic OR similarity)) AND (Title: (``natural language processing'' OR NLP OR ``Generative Artificial Intelligence'' OR ``GenAI'' OR ``Gen AI'') OR Abstract: (``natural language processing'' OR NLP OR ``Generative Artificial Intelligence'' OR ``GenAI'' OR ``Gen AI'')) AND  (Title:(Education) OR Abstract:(Education))) & With Educational Domain Constraint & 48 \\
         \hline
         IEEE & (((``semantic'' OR ``similarity''):Abstract OR (``semantic'' OR ``similarity''):``Document Title'') AND ((``natural language processing'' OR NLP OR ``Generative Artificial Intelligence'' OR GenAI OR ``Gen AI''):Abstract OR (``natural language processing'' OR NLP  OR ``Generative Artificial Intelligence'' OR GenAI OR ``Gen AI''):``Document Title'')) & Without Education Domain Constraint  & 362 \\
         \hline
         ACM & ((Title:(semantic OR similarity) OR Abstract:(semantic OR similarity)) AND (Title: (``natural language processing'' OR NLP OR ``Generative Artificial Intelligence'' OR ``GenAI'' OR ``Gen AI'') OR Abstract: (``natural language processing'' OR NLP OR ``Generative Artificial Intelligence'' OR ``GenAI'' OR ``Gen AI''))) & Without Education Domain Constraint & 869 \\
         \hline
    \end{tabular}
\end{table*}

\subsection{Phase-2: Pre-Screening}
This is the new phase introduced by L-PRISMA. As discussed previously, we want to filter out the relevant records for the literature survey during this phase. This phase uses the titles and abstract as the records identifiers. The first task is to define a statement which best describes the intent of the literature survey. For this we defined the intent and asked different GenAI tools (ChatGPT, Gemini and Claude) to refine our statement. Finally, we emerged with the following survey statement: \textit{``This investigates methods for measuring textual and semantic similarity between student-generated responses and reference answers, with a focus on applications in automated grading and educational assessment using natural language processing techniques.''} 

Note that, using GenAI to conceive such statement is not mandatory, however it can be helpful especially for non-native English speakers. It is also interesting to note that this search statement includes the complete context that the researcher is interested in. Because this helps the GenAI tools to better understand their intent.  

Secondly, We decided to use the Sentence-Bidirectional Encoder Representations from Transformers (S-BERT) model to fine tune the sentence pairs followed by the cosine similarity to find the semantic similarity between the survey statement and the records
~\cite{elmassry2025systematic}. 
S-BERT transforms sentences into vectors where semantic similarity can be measured efficiently, making it highly useful for NLP tasks that require understanding meaning across sentences or documents. This is also because by-far this is one of the best GenAI models to capture the sentence similarity, however, the researcher can use any other similarity model and the measurement matrix as well.
\begin{figure}
    \centering
    \includegraphics[scale=0.6]{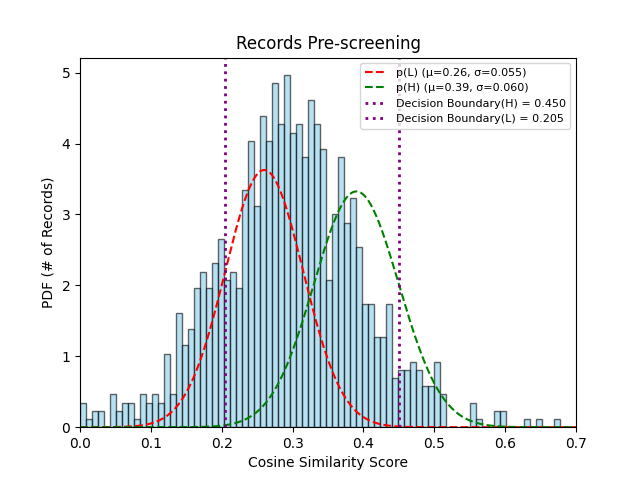}
    \caption{Probability Density Function (PDF) for number of records.}
    \label{fig:distribution}
\end{figure}

The Probability Density Function (PDF) of similarity scores for these records is plotted in Fig.~\ref{fig:distribution}. This can be clearly inferred from this graph that the similarity score of search outcome follows the Gaussian Mixture Model (GMM). Subsequently, in line with the Eq.~(\ref{eq:mixture-scores}), its easy to infer that this GMM can be considered a combination of two Gaussian distributions as described in Eq.~(\ref{eq:gmm})~\cite{stephen2007}. 

\begin{equation}
\begin{aligned}
p(s) = \sum_{k=H,L} \pi_k \, \mathcal{N}(s_i \mid \mu_k, \sigma_k^2), 
\end{aligned}
\label{eq:gmm}
\end{equation}

where \\
  \indent $s_i$ is the $i^{th}$ data point (i.e., cosine similarity score), \\
  \indent $\pi_k$ is the mixture weight, with $\sum_j \pi_j = 1$, \\
  \indent $\mu_k$, $\sigma_k^2$ are the mean and variance of component $k$, \\
  \indent $\mathcal{N}(x_i \mid \mu_k, \sigma_k^2)$ is the Gaussian density evaluated at $x_i$ under component $k$.

We obtained the distribution of these two Gaussian distributions using python GMM library. Since both of them are Gaussian distributed ($p_k(s) \sim \mathcal{N}(\mu_k, \sigma_k^2)$), we obtained the decision boundaries using the 2$\sigma_k$ bounds. 
The outcome of this has been plotted in the Fig.~\ref{fig:distribution}. At this stage the records below the lower cutoff points are removed (n=182). 

The records above the upper cutoff (n=60) are the most relevant records and needs to be manually screened during the screening phase while the records between the two decision boundaries can be considered to be screened and summarised by GenAI (n=989). It must be further noted that these are the guideline numbers and researchers can further tweak these numbers with their domain expertise.

\subsection{Phase-3: Screening}
This phase follows the standard PRISMA screening methodology with the following modifications. Records identified through the pre-screening phase undergo screening for eligibility assessment; Records (n=60) with high semantic similarity scores (above the statistical threshold determined in pre-screening) undergo traditional manual screening by human reviewers while Records (n=989) with lower semantic similarity scores undergo GenAI-assisted screening using structured prompts aligned with the eligibility criteria. For full-text retrieval, reports are obtained for both manual evaluation (high-relevance records) and GenAI-assisted evaluation (low-relevance records). Any reports that cannot be retrieved are documented. Subsequently, the reports among the high similarity scores are assessed manually and an excluded reports are recorded with justification for exclusion. It is important to record the specific LLM and the prompt being used during the entire process.

\subsection{Phase-4: Included}
This phase reports the included studies for the review and the number of reports in those studies. The only change here the studies are reported both for manual review by Humans as well as the GenAI reviews. The GenAI reviews are performed by the LLMs and the details of the LLMs are required to be be provided.

\section{Conclusions and Future Directions}\label{sec:conclusions}

In conclusion, this study introduces L-PRISMA, an enhanced adaptation of the PRISMA framework that incorporates recent advancements in GenAI for systematic literature reviews, thereby improving methodological efficiency. To address the inherent non-determinism of LLMs, the study further integrates human-led synthesis with a GenAI-assisted statistical pre-screening process. To elaborate the framework steps, an use case is also presented with corresponding statistical analysis. Future research should apply the process for different domains and literature to further refine the process. Moreover, further investigations are necessary to adapt to emerging GenAI capabilities and to develop domain-specific approaches that enhance the reliability and expand the applicability of the L-PRISMA framework.

\balance
\bibliographystyle{IEEEtran}
\bibliography{Myref}

\vspace{12pt}

\end{document}